# Totally Secure Classical Communication Utilizing Johnson (-like) Noise and Kirchoff's Law[1]

Laszlo B. Kish[2]

Texas A&M University, Department of Electrical Engineering, College Station, TX 77843-3128, USA; email: Laszlo.Kish@ee.tamu.edu

**Abstract**. An absolutely secure, fast, inexpensive, robust, maintenance-free and low-power-consumption communication is proposed. The states of the information bit are represented by two resistance values. The sender and the receiver have such resistors available and they randomly select and connect one of them to the channel at the beginning of each clock period. The thermal noise voltage and current can be observed but Kirchoff's law provides only a second-order equation. A secure bit is communicated when the actual resistance values at the sender's side and the receiver's side differ. Then the second order equation yields the two resistance values but the eavesdropper is unable to determine the actual locations of the resistors and to find out the state of the sender's bit. The receiver knows that the sender has the inverse of his bit, similarly to quantum entanglement. The eavesdropper can decode the message if, for each bits, she inject current in the wire and measures the voltage change and the current changes in the two directions. However, in this way she gets discovered by the very first bit she decodes. Instead of thermal noise, proper external noise generators should be used when the communication is *not* aimed to be stealth.

---

# 1. Introduction

Much effort has been spent to find a data communication methods which is absolutely secure. Quantum key distribution offers methods where the eavesdropper gets discovered as soon as she extracts a certain number of (noisy) bits (the number depends on the specific method), because of the extra noise she is introducing into the communication channel. Thus quantum encryption is considered to be absolutely secure. However they are expensive, slow, maintenance-needy and work only for short distances without a repeater.

Concerning classical physical information, there is a widespread opinion that classical information channels cannot be made secure. Such an opinion sounds reasonable because neither the *collapse of wavefunction during measurement* nor the *quantum no-cloning theorem*, which make quantum encryption secure, hold for classical information. Concerning today's classical encryptions, the weak point of the mathematical methods used generally is that the sender and receiver have to somehow share the key they shall use to encode and decode the message, while the eavesdropper is recording the communication between them. To share the key, they start with a common public information manipulated mathematically in a successive way, and send it back and forth several times until they are able to extract the key. The problem with this approach is that the eavesdropper may record everything and can also extract the key though she is only a passive observer with less information but a sufficient computational power at hand, such as a fictional quantum computer for factoring. Therefore those encryption methods, such as RSA encryption systems [1,2], where the sender and receiver *share some public information about the key*, are not absolutely secure.

The cipher to be described in this paper provides absolute security with classical information and it is based on the simultaneous physical encrypting of the information by both the sender and the receiver. It is relevant to mention that we have made an earlier mathematical attempt to double encrypt the message while bouncing it between the sender and receiver, the mathematical operator version of a *double-padlock* method [3]. However, finding of proper encryption (padlock) operators satisfying the necessary mathematical conditions of working/security is hopeless [2] with classical information, unless we use operators like those of RSA encryption systems [1], where there is a common public knowledge about the key that is however terminates the absolute security of the cipher. On the other hand, quantum operators can be applied easily and the result is a quantum communication scheme less noisy and more secure than earlier quantum key distribution systems [4]. However, even at this improved quantum case, the eavesdropper can extract many noisy bits until she will be discovered with a reasonable probability [5].

Another attempt [6] with classical (and quantum) information is the *stealth communication*. This is a zero-signal-power communication method utilizing parametric modulation of the classical or quantum noise in the information channel and it can provide absolute security for classical information with radio waves however only in thermal equilibrium [6]. If the eavesdropper uses cooled devices, she can hide for a much longer time period than one bit.

In this paper a new classical physical method is shown to make absolutely secure data transfer with classical information where the information is encrypted by both the sender and the receiver similarly to the double-padlock systems [3,4]. However the way of communication is very different from the double-padlock system, and it is very special because these encryptions take place simultaneously, and because the information is carried by *two physical quantities* with *zero crosscorrelation* between them: *voltage* and *current*. The method succeeds to achieve the difficult goal that there is no public knowledge about the keys, which are independent random sequences of the numbers 0 and 1 and independent Gaussian voltage noises generated locally at the two sides. They are not shared or transferred to anywhere and they are not saved, either. These properties makes the communication absolute secure.

While this communication is also absolutely secure like quantum communication, it has some major advantages: this communication is inexpensive to set up; very robust; maintenance-free; the eavesdropper is recognized after she extracts a single bit; it works for much longer distances without the need of repeaters; or if it is used for the same distance as the range of quantum communication, it is much faster. Because the eavesdropper has zero information, *computational power does not help to break the code* thus even an *infinitely advanced quantum computer would not be able to break this code*.

**2. Thermal noise voltage and current of two parallel resistors**

For the sake of readers unfamiliar with thermal noise in resistors, here we present some simple considerations and facts [7]. In thermal equilibrium at temperature $T$, in the low-frequency limit, the power density spectrum of the thermal noise voltage on an open-ended resistor $R$ is given by the

Johnson formula:

$$S_{u,th}(f) = 4kTR \qquad (1)$$

If the two ends of the resistor are shorted with a wire then the current in this loop will be

$$S_{i,th}(f) = \frac{4kT}{R} \qquad (2)$$

due to Ohm's law. Let us suppose that we have two resistors, $R_0$ and $R_1$, which are *connected parallel*. Then from Eq. 1 and the Kirchoff's laws it follows that the power density spectrum of the resultant noise voltage on the two parallel resistors will be:

$$S_{u,R\|}(f) = 4kT \frac{R_0 R_1}{R_0 + R_1} \quad . \qquad (3)$$

From Eq. 2, it follows that the power density spectrum of the noise current flowing through the loop consisting of the two resistors is:

$$S_{i,R\|}(f) = \frac{4kT}{R_0 + R_1} \qquad (4)$$

It is an important consequence of thermal equilibrium that the mean power flow between the parallel resistors is zero, the crosscorrelation between the voltage on these parallel resistors and the current flowing in this Kirchoff loop is zero.

## 3. Secure, threefold-encrypted Kirchoff-loop-Johnson-noise (KLJN) cipher

### 3.1 Physical base

Let us suppose that both the sender and the receiver have a resistor each and these resistors are connected parallel by a pair of wires, which represents the information channel between the sender and the receiver. The eavesdropper has no information about the resistance of these resistors. The key questions are:

i) Can the eavesdropper find out the resistance values from voltage and current measurements?

ii) If yes, can she find out the location of the resistances; which resistor is at the sender and which one is at the receiver?

The answer to question i) is *yes*. By measuring the current in the wire and the voltage between the wires, the eavesdropper can determine $S_{u,R\|}(f)$ and $S_{i,R\|}(f)$. Then using Eqs. 3,4 she can set up a second order equation and the two solutions provide the two resistance values. However, the answer to question ii) is *no* and this answer guarantees the absolute security of the KLJN cipher. There is no way to determine the exact location of these resistances. The specific scaling between the resistance and the thermal noise voltage and current (see Eqs. 1,2) makes it sure that the net energy flow between the sender and the receiver is zero, thus even the Pointing vector or similar cross-correlation measures provide zero information about the location.

## 3.2 The absolutely secure communication scheme

Accordingly, the absolute secure classical communication scheme utilizing Kirchoff laws and Johnson noise is shown in Fig. 1. The information channel is a wire. The sent information bit is carried by the sender's choice of resistor $R_s$ selected from two significantly different values, $R_0 << R_1$, where the two values represent the two states of the information bit (0 and 1). The sender encrypts this message with the random generator voltage $U_S(t)$ of the actual resistor. The receiver double-encrypts the message by using his randomly chosen resistor $R_R$, which (for simplicity but not necessarily) can be selected from the same set of two resistance values and the corresponding random generator voltage $U_R(t)$. At each clock period, a new information bit is sent and, at the same time, the receiver's resistance is changing randomly between the two values. The clock period is selected so that a sufficiently good statistics can be made on the measured noise voltage or current in the channel. The random voltage generators are either the Johnson noises of the resistors or they are artificial noise generator with much larger noise voltage but with the same linear scaling relation between the resistance and the noise voltage spectrum as that of Johnson noise. We call this a *threefold encryption scheme* because the data are encrypted by three independent stochastic processes: the two noise voltage generators, and the resistance at the receiver. Because the sender and receiver know their own actual resistor value they can both determine the resistor value at the other side by measuring either the voltage or the current using Eq. 3 or 4, respectively. For the identification of *secure bits*, see below, it is *necessary* that the sender also determines the resistance value at the other side and that can be used for a simultaneous two-way communication.

The eavesdropper may also have access to the measurement of the voltage and current in the channel, see Figure 2, however this information is not enough to break the code. As we have shown it above, by measuring both the voltage and current in the channel, she can extract only the two momentary resistance values however she is unable to locate the resistors when their resistances are different, consequently she cannot decide if the sender is communicating 0 or 1.

**3.3. Non-secure bits and simultaneous two-way communication.**

If both the sender and receiver use the same resistance value, the eavesdropper can recognize that fact because the resultant resistance is either the lowest or the highest of the three possibilities, thus she can find out the resistor value and the status of the information bit at the sender. Thus such a bit is not secure and cannot be used for secure data transfer. On the average, the 50% of the sender and receiver bits will be different thus we can use only that 50% for secure communication. Similar situations are well known in quantum communication and it is easy to make use of the successfully communicated secure bits because the sender is aware of the security or non-security of the communicated bit. For example, after a whole random sequence is sent, it is possible to announce through a public channel which secure bits forms the message and what is their proper order. In this way, both the sender and the receiver can simultaneously communicate a random message and the random message will be the inverse of each other at the two sides, similarly to *quantum-entanglement* in quantum information processing. Thus also the receiver can send information, simultaneously; therefore *both sides can simultaneously act as sender and receiver* by randomly choosing between the two resistor values at each clock period. In this way the 50% bit loss due to the security requirement is made up by the simultaneous operation mode which is a speed doubler.

## 3.4 External noise generators

We have already mentioned above that proper external noise generators can and should also be used because the same considerations hold for resistances and corresponding external noise voltage generators that follow the same scaling law between the resistance and the mean-square voltage noise as that of the thermal noise:

$$S_{u,th}(f) = VR \qquad , \tag{5}$$

where $V$ is the scaling coefficient and its value has to be equal at the senders and the receiver's sides. Then all considerations and conclusions described above remain valid including the absolute security. The advantage of using external noise generators is that $V$ can be arbitrarily large instead of the $4kT$ value resulting in the case of thermal noise. Thus, with using external noise generator of large $V$ makes the communication free of background noise effects, which is a strong advantage compared to quantum communication where the background noise of the quantum source, the channel and the detectors is significantly reducing performance and compromising security. The only advantage of the thermal noise driving, which has originally been shown for the sake of simplicity, is that it is a *stealth communication* method using only the background noise in the channel, see [6]. In this respect, Figure 1 shows also the way to provide *absolute security for the stealth communication* arrangement with wire connection that has not been achieved in the original proposal of secure communication [6].

## 4. Breaking the code and the immediate detection of the eavesdropper

It is possible to physically break the KLJN cipher, see Fig. 3, if the eavesdropper injects a small stochastic current $\Delta I_E(t)$ into the channel and determines the following crosscorrelation functions:

$$C_{ES} = \langle \Delta I_E(t) \Delta I_{ES}(t) \rangle \qquad \text{and} \qquad C_{ER} = \langle \Delta I_E(t) \Delta I_{ER}(t) \rangle \;, \tag{6}$$

where $\Delta I_{ES}(t)$ and $\Delta I_{ER}(t)$ are the induced current changes by the induced voltage change $\Delta U_{E,Ch}(t)$ and the linear response. Obviously, the following rule due to Kirchoff's law allows the determination of the information bit:

$$R_R / R_S = C_{ES} / C_{ER} \ , \tag{7}$$

where $R_S$ and $R_R$ are the actual resistances at the sender and the receiver, respectively. The easiest way to uncover the eavesdropper is to use a public channel to exchange and compare the measured current values at the sender and the receiver. In this case, the eavesdropper is *immediately discovered, before she can make any statistics*, because she needs to cause a measurable current difference while the entangled bits are communicated. Thus the eavesdropper has only one choice: instead using small current and making a statistics, to inject a fast and large current spike in the channel and use the relation

$$R_R / R_S = \Delta I_{ES} / \Delta I_{ER} \tag{8}$$

to extract the bit values. Thus the eavesdropper can extract a single bit but she gets immediately uncovered. Therefore the communication is absolutely secure. It is more secure than quantum communication where a certain number of noisy bits can always be extracted by a quantum amplifier before the eavesdropper is uncovered.

## 5. Practical questions: speed, range and cost

There are two factors limiting speed and range. Firstly, Kirchoff's laws should hold. That means that

$$L << \lambda_{max} = c / f_{max} \qquad \text{or} \qquad f_{max} L << c \ , \tag{9}$$

where $L$, $f_{max}$ and $c$ are the range, the maximal frequency bandwidth and the velocity of EM waves in the wire, respectively. Secondly, the *clock frequency* (*effective bandwidth*) $f_c$ should be low-enough for a reasonably good noise statistics to distinguish the resistors $R_0 << R_1$ securely, thus $f_c << f_{max}$. For this reason, it is much faster to use the measurement of the mean-square noise voltage and current which are the integral of the corresponding power density spectra [7] and provide a good statistics in a much shorter time.

For a practical estimation, let us suppose that $c = 2*10^8$ meter/s, $f_{max}L = 0.1*c$, and $f_c = 0.1*f_{max}$. Then the effective *bandwidth-distance product* $f_cL = 2*10^6$ meterHz. This is slightly (factor of 2-3) better than present quantum communicator arrangements [8].

However, there is a simple and inexpensive way to improve speed radically. Because this communicator is extremely inexpensive, 100-1000 times cheaper than quantum communicators of the similar performance, it is natural to build parallel channels to enhance speed and security. Using chip technology and a multi-wire cable with 100 wires, the speed of the system can be increased by a factor of 100 and the price would still be much below the price of a quantum communicator. Moreover, the security can further be enhanced because the information can be cross-encrypted among the wires by a public code thus, if the eavesdropper would read out a single bit from a single wire, that would still be zero bit of extracted information without reading the other 99 wires.

Finally it should be mentioned that this technology is not only inexpensive but also very robust and virtually maintenance-free compared to quantum communicators, moreover it consumes much lower power. It is very robust: vibrations, dust, temperature gradients and material ageing, which are serious problems in quantum communication, cause no problem here.

## 6. Summary

In this Letter, we outlined a new way of communication through a wire. The main advantages of this method are absolute security and the use of classical information in the form that is fast, inexpensive, robust (vibration, dust, temperature and aging resistant), maintenance free and it has low power consumption. Even quantum computers are unable to break this code moreover it is more secure than quantum communication. If it is applied for the same range as a quantum communicator, its speed can easily be made 100 times greater; or alternatively, at the same speed it can go for a 100 times longer distance without the need of repeater stations. The communication can also use stealth mode when the noise generators are the thermal noises of the resistors.

**Acknowledgements**

**Figure caption**

**Figure 1.**
Absolute secure classical communication scheme utilizing Kirchoff laws and a threefold encryption. The information channel is a wire. The message is carried by the sender's choice of resistor value. The sender encrypts this message by the random generator voltage $U_S$. The receiver double-encrypts the message by using his randomly chosen resistor $R_R$ and the random generator voltage $U_R$. The random voltage generators are either the Johnson noises of the resistors or artificial noise generator with much larger noise voltage but with the same scaling relation between the resistance and the noise voltage spectrum as that of Johnson noise. The eavesdropper may have access to the measurement of the voltage and current in the channel, however this information is not enough to break the code.

**Figure 2.**
Circuit models for the calculation of the channel voltage and current via the Kirchoff's law and the Johnson noise equation. The eavesdropper can measure the voltage and the current and, after determining the spectra she can derive the resistor values from Eqs. 3 and 4, but not their location.

**Figure 3.**
Breaking into the system needs to inject a current in the channel. The current has to have different dynamical and statistical characteristics than the voltage noises used by the sender and the receiver. However, as soon as the eavesdropper extracts a single bit her action gets discovered because the sender and the received observe the same deviations in the voltage, the current and their statistical properties.

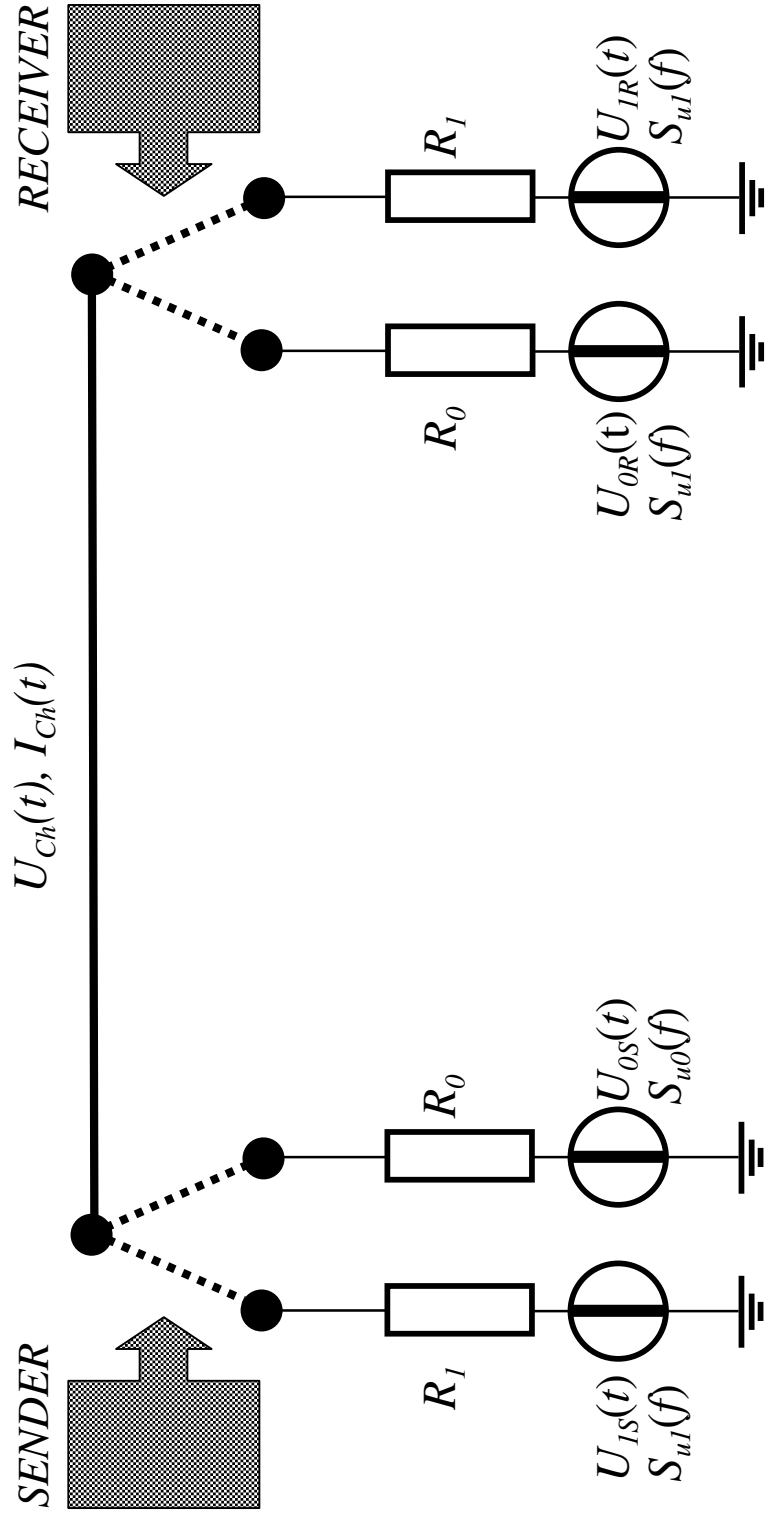

**Figure 1.**

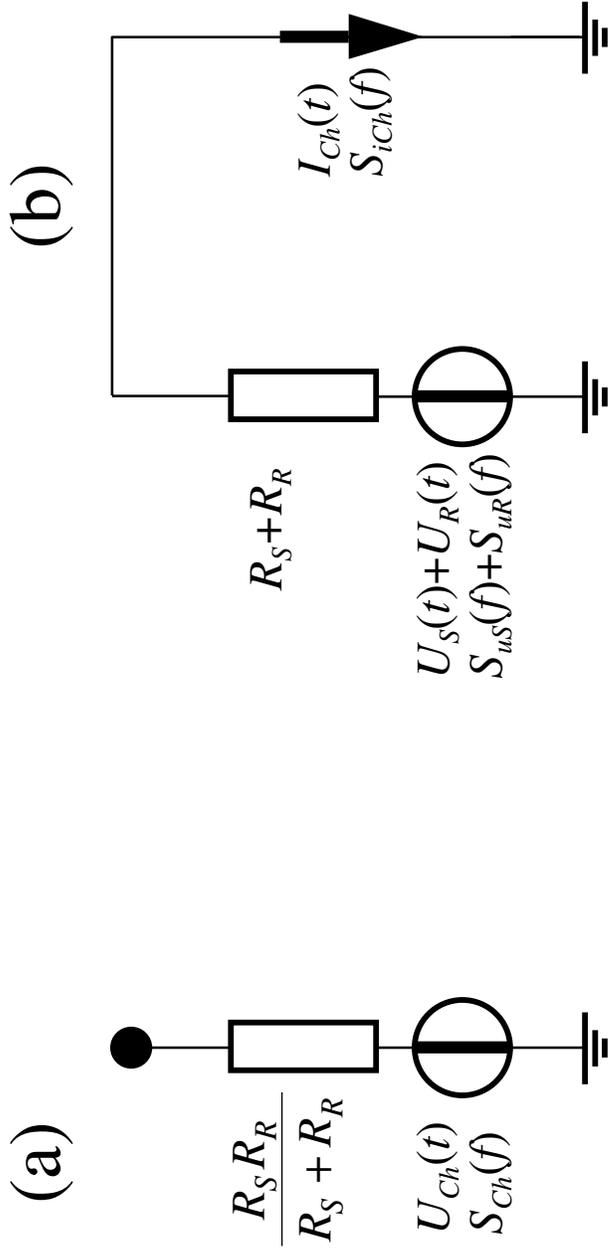

**Figure 2.**

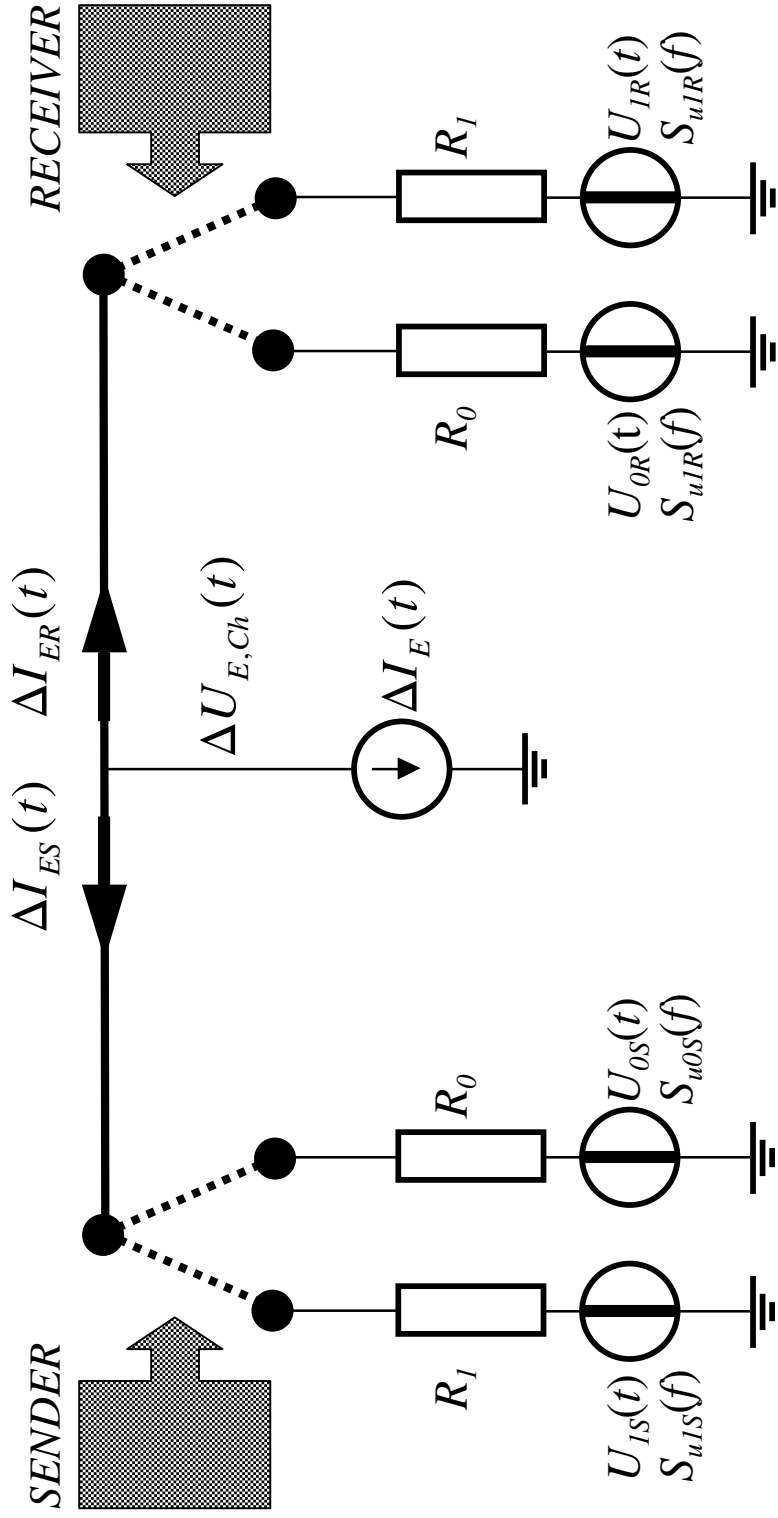

**Figure 3.**